\documentstyle[12pt]{article}
\title{\bf Fluctuating Strings\\
in the Universal Confining\\ 
String Theory and Gluodynamics}
\author{D.V.ANTONOV \thanks{ E-mail addresses:
antonov@pha2.physik.hu-berlin.de, antonov@vxitep.itep.ru}{\,}
\thanks {Supported
by Graduiertenkolleg {\it Elementarteilchenphysik}, Russian
Fundamental Research Foundation, Grant No.96-02-19184, DFG-RFFI,
Grant 436 RUS 113/309/0, and by the INTAS, Grant No.94-2851.}
\\
{\it Institute of Theoretical and Experimental Physics,}\\
{\it B.Cheremushkinskaya 25, 117 218, Moscow, Russia}\\
{\it and}\\
{\it Institut f\"ur Physik, Humboldt-Universit\"at,}\\
{\it Invalidenstrasse 110, D-10115, Berlin, Germany}}
\date{}
\begin{document}
\maketitle
\vspace{1mm}
\centerline{\bf {Abstract}}
\vspace{3mm}
The effective string theory emerging from the bilocal approximation to 
the Method of Vacuum 
Correlators in gluodynamics  
is shown to be well described by the 4D theory of the massive Abelian  
Kalb-Ramond field interacting with the string, which is known to 
be the low-energy limit of the Universal Confining 
String Theory. This correspondence follows from the agreement 
of the behaviour of the coefficient functions, which parametrize 
the gauge-invariant correlator of two gluonic field strength tensors, 
known from the lattice data, with their values obtained from the 
propagator of the Kalb-Ramond field. We discuss this correspondence 
in several aspects and demonstrate that the mass of the Kalb-Ramond 
field in this approach plays the role of the inverse correlation 
length of the vacuum, so that in the massless limit string picture 
disappears. Next, we apply the background field method, known in the 
theory 
of nonlinear sigma models, to obtain the action, which is quadratic in  
quantum fluctuations around a given (e.g. minimal) 
string world-sheet. Several nontrivial types of couplings of 
these fluctuations with the background world-sheet are obtained and 
discussed.        

\vspace{6mm}
{\large \bf 1. Introduction}

\vspace{3mm}

Recently, a new approach to the string representation of the confining 
phase of gauge theories was proposed$^{1}$. This is the so-called 
Universal Confining String Theory (UCST), which is the theory of 
an Abelian antisymmetric tensor field with a nonlinear action, interacting 
with the string. This field effectively substitutes an 
infinite number of monopoles in the 3D compact QED. It was argued in 
Ref. 1 that the summation over the branches of the UCST action should 
correspond to the summation over the string world-sheets. It was also 
proved in Ref. 1 that the UCST partition function, which is nothing else 
but the Wilson average in the 3D compact QED, satisfies loop 
equations$^{2}$ modulo contact terms. However, the statement made 
in Ref. 1 concerning the universality of the wave operator standing 
on the L.H.S. of the usual loop equations obtained from the Yang-Mills 
theory, which were derived and 
investigated in Ref. 2, was only a conjecture, and the relation of 
these equations to the equation of motion of the tensor field in the 
UCST was absolutely unclear. In order to clarify this conjecture, 
in Ref. 3 the loop equation for the 4D UCST partition function 
was derived and investigated. In particular, it was demonstrated in$^{3}$ 
that the wave operator of the obtained loop equation is 
quite different from the usual one. Also the corresponding contact term 
was calculated explicitly. 

The low-energy limit of the UCST, in which the Wilson loop could be 
evaluated exactly, was discussed in Ref. 1 and investigated in the 4D case 
in Ref. 4, 
where the string tension of the Nambu-Goto term and the coupling 
constant of the rigidity term were calculated. The latter one occured to be  
negative, which means that the obtained (Euclidean) 
string effective action is stable$^{5}$. Also the 4D UCST action was 
derived in Ref. 4 by performing the exact duality transformation, 
while in Ref. 1 it was done only at the semiclassical level.      

However, as it was first already pointed out in Ref. 6, the non-Abelian 
generalization of the above described antisymmetric tensor theory 
is quite difficult. 

Another, more phenomenological, but on the other hand adapted to  
the non-Abelian case approach to the problem of the string representation 
of the confining phase of gauge theories was developed in Ref. 7-9. It is 
based on the Method of Vacuum Correlators$^{10}$ (MVC) and allows one to get 
the information about the gluodynamics string effective action from 
the expansion of the Wilson 
average, considered as a statistical weight of this string theory, in powers 
of the correlation length of the vacuum. In this way, in Ref. 7 the 
sign of the rigidity term was found to be negative according to the present 
lattice data, which means that the MVC is 
in agreement with the dual superconductor model of confinement due to
Ref. 5. In Ref. 8, the string effective action obtained in Ref. 7, was 
applied to the derivation of the correction to the Hamiltonian of the 
QCD string with quarks$^{11}$ due to the rigidity term. However, up to 
now it is not clear how the summation over the world-sheets could appear 
within this approach. The most appropriate way in this direction might 
lie in the accounting for the perturbative gluons, which propagate inside 
the Wilson loop and, as it was argued in Ref. 12, should generate 
world-sheet excitations. In Ref. 9, this conjecture was elaborated out 
by virtue of the integration over perturbative gluons in the Wilson average, 
which in the lowest order of perturbation theory leads to the interaction 
of the world-sheet elements via the exchanges of perturbative gluons, 
propagating in the nonperturbative background. Expanding this interaction 
in powers of the derivatives w.r.t. the world-sheet coordinates, we finally 
get in the lowest order of this so-called curvature expansion some 
definite correction to the rigid string coupling constant, while the 
string tension of the Nambu-Goto term acquires no additional contribution 
due to perturbative gluons and keeps its pure nonperturbative value.

In this Letter, we shall propose an alternative method of description 
of fluctuations of the gluodynamics string world-sheet. To this end, we 
shall first demonstrate in the next Section that the gluodynamics 
string emerging 
within the bilocal approximation to the MVC  
could be effectively described by the same action of the massive Kalb-Ramond 
field interacting with the string, which describes the low-energy limit 
of the UCST. This observation will be proved by the calculation of the 
coefficient functions standing at two Kronecker structures 
(which describe surface and 
boundary terms in the string effective action) in the propagator of 
the Kalb-Ramond field, and further comparison of 
them with the values of the coefficient functions standing at the 
corresponding structures in the gauge-invariant correlator of two 
gluonic field strength tensors. The latter ones are unfortunately yet  
not found 
from the gluodynamics Lagrangian, while equations for correlators 
were obtained in Refs. 13 and 14 and then investigated in Ref. 15 
by making use 
of the stochastic quantization method (see also Ref. 16, where alternative  
equations for 
correlators following from the non-Abelian Bianchi 
identities, which were proposed in Ref. 17 and generalized in Ref. 13, 
were investigated). That is why, we shall compare the coefficient functions, 
standing in the propagator of the Kalb-Ramond field, with the lattice 
data$^{18}$ concerning the behaviour of the corresponding coefficient 
functions in gluodynamics 
(see also Ref. 19, where these functions were measured in QCD with 
dynamical fermions). We shall see that the mass of the Kalb-Ramond field 
plays the role of the inverse correlation length of the vacuum, so that  
in the sum rules' limit, when the correlation length of the vacuum tends 
to infinity (see the last Ref. in 10), which in our model corresponds 
to the case of the 
massless Kalb-Ramond field, 
the string picture disappears, and we are left with 
the boundary terms only. 
The points described above will be the topics of the next Section.

In Section 3, we shall use the model of the gluodynamics string proposed 
in Section 2, to describe string world-sheet excitations. Correspondingly, 
for the low-energy limit of the UCST this will be an exact procedure rather 
than a model dependent approach. The world-sheet excitations will be 
described with the help of the background field method developed in 
Ref. 20 
for nonlinear sigma models. Namely, we shall split the world-sheet 
coordinate into a background and quantum fluctuations, after which upon the 
integration over the Kalb-Ramond field we shall derive for 
the latter ones a quadratic action, which contains several nontrivial 
couplings of quantum fluctuations 
with the background world-sheet with- and without derivatives.
 
The main results of the Letter are summarized in the Conclusion.

In the Appendix, we perform rather nontrivial integration over the 
Kalb-Ramond field in the expression for the 4D UCST partition function.

\vspace{6mm}

{\large \bf 2. A Unified Description of the Gluodynamics String and 
the Low-Energy Limit of the UCST}

\vspace{3mm}
The partition function of the 4D UCST (which is nothing else but the 
Wilson average in the Euclidean 4D compact QED) in the low-energy limit 
has the form$^{1,4}$

$$\left<W(C)\right>=N\int DB_{\mu\nu}\exp\Biggl[
\int dx\Biggl(-\frac{1}
{12\Lambda^2}
H_{\mu\nu\lambda}^2-\frac{1}{4e^2}B_{\mu\nu}^2+iB_{\mu\nu}T_{\mu\nu}
\Biggr)\Biggr]. \eqno (1)$$ 
Here 

$$H_{\mu\nu\lambda}=\partial_\mu B_{\nu\lambda}+\partial_\nu 
B_{\lambda\mu}+\partial_\lambda B_{\mu\nu}$$ 
is a strength tensor of the 
field $B_{\mu\nu}$, 

$$T_{\mu\nu}(x)=\int d\sigma_{\mu\nu}(x(\xi))\delta (x-x(\xi))$$ 
is the vorticity 
tensor current, $e$ is a dimensionless coupling constant, and $\Lambda
\equiv\frac{\Lambda_0}{4}\sqrt{z}$, where $z\sim {\rm e}^{-\frac{\rm const.}
{e^2}}$, and $\Lambda_0$ is a cutoff which is necessary in 
4D. Gaussian integration in (1) is carried out in the Appendix and 
yields the UCST low-energy action in the 
form  

$$S_{{\rm UCST}}=\int d\sigma_{\lambda\nu}(x)\int d\sigma_{\mu\rho}(x')
\left<B_{\lambda\nu}(x)B_{\mu\rho}(x')\right>. \eqno (2)$$
Here

$$\left<B_{\lambda\nu}(x)B_{\mu\rho}(0)\right>\equiv\left<B_{\lambda
\nu}(x)B_{\mu\rho}(0)\right>^{(1)}+\left<B_{\lambda\nu}(x)B_{\mu
\rho}(0)\right>^{(2)}, \eqno (3)$$
where

$$\left< B_{\lambda\nu}(x)B_{\mu\rho}(0)\right>^{(1)}=
\frac{e^2m^3}{8\pi^2}
\frac{K_1\left(m\left|x\right|\right)}{\left|x\right|}
\Biggl(\delta_{\lambda
\mu}\delta_{\nu\rho}-\delta_{\mu\nu}\delta_{\lambda\rho}\Biggr),\eqno (4)$$

$$\left<B_{\lambda\nu}(x)B_{\mu\rho}(0)\right>^{(2)}=\frac{e^2m}{4\pi^2
x^2}\Biggl[\Biggl[\frac{K_1\left(m\left|x\right|\right)}{\left|x\right|}+
\frac{m}{2}\Biggl(K_0\left(m\left|x\right|\right)+K_2\left(m\left|x\right|
\right)\Biggr)\Biggr]\Biggl(\delta_{\lambda\mu}\delta_{\nu\rho}-
\delta_{\mu\nu}\delta_{\lambda\rho}\Biggr)+$$

$$+\frac{1}{2\left|x\right|}\Biggl[3\Biggl(\frac{m^2}{4}+\frac{1}{x^2}\Biggr)
K_1\left(m\left|x\right|\right)+\frac{3m}{2\left|x\right|}\Biggl(K_0
\left(m\left|x\right|\right)+K_2\left(m\left|x\right|\right)\Biggr)+
\frac{m^2}{4}K_3\left(m\left|x\right|\right)\Biggr]\cdot$$

$$\cdot\Biggl(\delta_{\lambda
\rho}x_\mu x_\nu+\delta_{\mu\nu}x_\lambda x_\rho-\delta_{\mu\lambda}
x_\nu x_\rho-\delta_{\nu\rho}x_\mu x_\lambda\Biggr)\Biggr]. \eqno (5)$$
In Eqs. (4) and (5), $m\equiv\frac{\Lambda}{e}$ is the mass of the 
Kalb-Ramond field, $K_i$'s, $i=
0,1,2,3$, stand for the Macdonald functions, and one can show that the term  

$$\int d\sigma_{\lambda\nu}(x)\int d\sigma_{\mu\rho}(x')\left<
B_{\lambda\nu}(x)B_{\mu\rho}(x')\right>^{(2)}$$
could be rewritten as a boundary one, due to which one can immediately 
establish a correspondence between $\left<W(C)\right>=\exp\left(-
S_{\rm UCST}\right)$ and the Wilson average 
written within the bilocal approximation to the MVC. This correspondence 
yields the following values of the coefficient functions $D$ and $D_1$, 
which parametrize the gauge-invariant correlator of two gluonic field 
strength tensors

$$D\left(m^2x^2\right)=\frac{e^2m^3}{8\pi^2}\frac{K_1(m\left|x\right|)}
{\left|x\right|} \eqno (6)$$
and

$$D_1\left(m^2x^2\right)=\frac{e^2m}{4\pi^2x^2}\Biggl[\frac{K_1(m
\left|x\right|)}{\left|x\right|}+\frac{m}{2}\Biggl(K_0(m\left|x
\right|)+K_2(m\left|x\right|)\Biggr)\Biggr]. \eqno (7)$$

Notice, that the derivative (or curvature) expansion of the action (2), 
which in 
this case is 
equivalent to the $\frac{1}{m}$-expansion, was performed 
in Ref. 4 and up to a common constant positive factor has the form 

$$S_{{\rm UCST}}=m^2 K_0\left(\frac{\sqrt{z}}{4e}\right)\int d^2\xi\sqrt{g}-
\frac{1}{4}\int d^2\xi\sqrt{g}g^{ab}\left(\partial_at_{\mu\nu}\right)
\left(\partial_bt_{\mu\nu}\right)+\frac{m}{2\pi}f\left(\frac{\sqrt{z}}
{4e}\right)\int\limits_0^1 ds\sqrt{\frac{dx_\mu}{ds}\frac{dx_\mu}
{ds}}. \eqno (8)$$
Here $g_{ab}$ and $t_{\mu\nu}$ stand for the induced metric and the extrinsic 
curvature tensor of the string world-sheet respectively, 

$$f(y)\equiv\int\limits_y^{+\infty}\frac{dt}{t}K_1(t),$$		
$x_\mu(s)$ in the last term on the R.H.S. of Eq. (8) parametrizes the 
contour $C$ in Eq. (1), $x_\mu(0)=x_\mu(1)$, and we have omitted the 
full derivative term of the form $\int d^2\xi\sqrt{g}R$, where $R$ is 
a scalar curvature of the world-sheet. As it was already mentioned in 
the Introduction, the coupling constant of the rigidity term in this 
expansion is negative, which means the stability of the string.

Asymptotic behaviours of the functions $D$ and $D_1$ following from 
Eqs. (6) and (7) at $\left|x\right|\ll\frac1m$ and $\left|x\right|\gg
\frac1m$ read

$$D\longrightarrow\frac{e^2m^2}{8\pi^2x^2}, \eqno (9)$$

$$D_1\longrightarrow\frac{e^2}{2\pi^2\left(x^2\right)^2} \eqno (10)$$
and

$$D\longrightarrow \frac{e^2m^4}{8\sqrt{2}\pi^{\frac32}}\frac{{\rm e}^
{-m\left|x\right|}}{\left(m\left|x\right|\right)^{\frac32}}, \eqno (11)$$

$$D_1\longrightarrow\frac{e^2m^4}{4\sqrt{2}\pi^{\frac32}}\frac{{\rm e}^
{-m\left|x\right|}}{\left(m\left|x\right|\right)^{\frac52}} \eqno (12)$$
respectively. 

Let us comment on the asymptotic behaviours (9)-(12). Firstly, 
one can see that Eq. (10) is in agreement with the MVC, 
where due to Ref. 10 
at the distances which are much smaller than the correlation length 
of the vacuum $T_g$,

$$D_1\longrightarrow \frac{16\alpha_s\left(x^2\right)}
{3\pi\left(x^2\right)^2}. \eqno (13)$$
The difference in the numerical constants in the asymptotic 
behaviours (10) and (13) 
is due to the colour factor in the one-gluon exchange diagram, which 
contributes into Eq. (13). This factor, which is absent in the asymptotics 
(10) of the Abelian propagator, could be accounted for by the proper 
tuning of the charge $e$ of the Kalb-Ramond field, if we replace the 
running 
$\alpha_s(x^2)$ in Eq. (13) by some fixed value, say approximate 
$\alpha_s(x^2)$ by its ``frozen'' value, which it acquires  
at the confinement scale$^{21}$ 
$\frac{1}{\sqrt{\sigma}}$, where $\sigma$ is the 
string tension of the Nambu-Goto term. 

However Eq. (9) 
agrees with MVC only in the 
lowest order of perturbation theory, when the perturbative part of 
the function $D$ is absent, and $D$ tends to the gluonic condensate 
at $\left|x\right|\to 0$, i.e. this agreement is only in a sense that 
$D\ll D_1$. One can see that Eq. (9) does not coincide 
with the behaviour 
of the function $D\left(x^2\right)$ of the type 

$$\frac{1}{\left(x^2\right)^2}\Biggl(\alpha\ln\left(M^2x^2\right)+
\beta\Biggr),$$
where $\alpha,\beta$ stand for some constants, and $M$ is a certain 
mass parameter, which takes place in the next-to-leading order 
of perturbation 
theory$^{22}$. This is due to the fact that such a behaviour is a 
specific property of non-Abelian theories$^{23}$ and presumably 
could not be reproduced by any Abelian model. 

Secondly, comparing asymptotic behaviours (11) and (12) we see that 
the function $D_1$ falls off much faster than the function $D$, which 
is in agreement with the lattice data$^{18}$. The exponential 
falls-off of the functions $D$ and $D_1$ also agree with Ref. 18, while 
the preexponential power-like behaviours do not. However the $\frac
{1}{\left(x^2\right)^2}$-fits of these behaviours used in Refs. 18 
and 19 were motivated by the short-distance asymptotics of the function 
$D_1$, which according to Eq. (10) is reproduced in our model as well. 

Eqs. (11) and (12) tell us once more that the mass $m$ of the Kalb-Ramond 
field 
in our approach should be treated as an inverse correlation length of the 
vacuum $T_g$, so that in the ``string limit of QCD''$^{24}$, when the 
string tension of the Nambu-Goto term, $\sigma$, 
is kept fixed at vanishing $T_g$, we have a correspondence of 
the type $m\sim
\sqrt{\frac{D(0)}{\sigma}}$. In the opposite regime when the Kalb-Ramond 
field is massless (or equivalently, in the strong coupling regime of the 
theory (1)), which corresponds to 
the QCD sum rules' case (see the last Ref. in 10), 
the ``confining'' part (4)  
of the propagator of the Kalb-Ramond field vanishes, and we are left only 
with the ``boundary'' part (5), whose contribution to $S_{{\rm UCST}}$ 
takes the form 

$$\int d\sigma_{\lambda\nu}(x)\int d\sigma_{\mu\rho}(x')\left<
B_{\lambda\nu}(x)B_{\mu\rho}(x')\right>^{(2)}=\frac{e^2}{2\pi^2}
\oint\limits_C^{}dx_\mu\oint\limits_C^{}dx'_\mu\frac{1}{(x-x')^2},$$
which could be anticipated from the very beginning. 
This result is also in agreement with Ref. 4, where it was shown that in the 
strong coupling regime of the theory (1) strings are completely 
suppressed when $\Lambda_0\to +\infty$.

\vspace{6mm}
{\large \bf 3. Description of the World-Sheet Excitations}

\vspace{3mm}
In order to describe fluctuations of strings in the model (1), we shall 
adopt the background-field method developed in Ref. 20 for the 
nonlinear sigma models. The three differences of our case with the one 
considered in Ref. 20 are the absence of the Polyakov term in the action, 
flatness of the field manifold (i.e. the space-time), 
since we consider an arbitrary Wilson 
loop, not necessarily lieing on the unit sphere, and the necessity of the 
eventual integration over the field $B_{\mu\nu}$. Consequently, the 
geodesics passing through the background manifold $y_\mu (\xi)$ and the 
excited manifold $x_\mu (\xi)=y_\mu (\xi)+z_\mu (\xi)$, where $z_\mu 
(\xi)$ stands for the world-sheet fluctuation, are straight,   
$\rho_\mu(\xi,s)=y_\mu (\xi) +sz_\mu (\xi)$, where $s$ denotes the 
arc-length parameter, $0\le s\le 1$. The expansion of the action (2) 
in powers of quantum fluctuations $z_\mu (\xi)$ could be performed 
by virtue of the following generating functional, which is the 
arc-dependent term describing the interaction of the Kalb-Ramond 
field with the string in 
Eq. (1),

$$I\left[\rho(\xi,s)\right]=i\int d^2\xi B_{\mu\nu}\left[\rho(\xi,s)
\right]\varepsilon^{ab}\left(\partial_a\rho_\mu (\xi,s)\right)\left(
\partial_b\rho_\nu (\xi,s)\right). \eqno (14)$$
Then the term containing $n$ quantum fluctuations reads as 

$$I^{(n)}=\left.\frac{1}{n!}\frac{d^n}{ds^n}I\left[\rho(\xi,s)\right]
\right|_{s=0},$$
and we get from Eq. (14)

$$I^{(0)}=i\int d\sigma_{\mu\nu}(y(\xi))B_{\mu\nu}\left[y(\xi)
\right],\eqno (15)$$

$$I^{(1)}=i\int d\sigma_{\mu\nu}(y(\xi))z_\lambda(\xi)H_{\mu\nu\lambda}
\left[y(\xi)\right], \eqno (16)$$
and

$$I^{(2)}=i\int d^2\xi z_\nu(\xi)\varepsilon^{ab}\left(\partial_a 
y_\mu(\xi)\right)\Biggl(\left(\partial_b z_\lambda(\xi)\right)
H_{\nu\mu\lambda}\left[y(\xi)\right]+\frac{1}{2}z_\alpha (\xi)\left(
\partial_b y_\lambda(\xi)\right)\partial_\alpha H_{\nu\mu\lambda}
\left[y(\xi)\right]\Biggr), \eqno (17)$$
where during the derivation of Eqs. (16) and (17) we have omitted several  
full derivative terms. 

It is worth mentioning, that as it was discussed in Ref. 20, 
the terms (15)-(17) are 
necessary and sufficient to determine all one-loop ultraviolet 
divergences in the theory (14) at $s=0$. This 
statement is important for the model of the gluodynamics string with 
the action (2), proposed in the previous Section, since in the  
UCST case, Eq. (2) describes only its low-energy effective action, which does 
not suffer from the ultraviolet divergences. It is also important for the 
Abelian Higgs Model in the Londons' limit$^{25}$, where one can 
show$^{26}$ that the partition function (1) with an additional 
integration over metrics and $T_{\mu\nu}(x)$ 
corresponding to a closed surface 
is nothing else but a 
't Hooft loop average defined on the string world-sheet.  

In order to get the desirable action, quadratic in quantum fluctuations 
$z_\mu(\xi)$, we shall first carry out the integral

$$N\int DB_{\mu\nu}\exp\Biggl[-\int dx\Biggl(\frac{1}{12\Lambda^2}
H_{\mu\nu\lambda}^2+\frac{1}{4e^2}B_{\mu\nu}^2\Biggr)+I^{(0)}+I^{(1)}
\Biggr]. $$
It occurs to be equal to 

$$\exp\Biggl[-\int d\sigma_{\lambda\nu}\left(y(\xi)\right)
\int d\sigma_{\mu\rho}\left(y(\xi')\right)
\Biggl(\left<B_{\lambda\nu}\left[y(\xi)\right]B_{\mu\rho}\left[y(\xi')
\right]\right>^{(1)}+2z_\alpha(\xi)\cdot$$

$$\cdot\frac{\partial}{\partial y_\alpha(\xi)}\left<
B_{\lambda\nu}\left[y(\xi)\right]B_{\mu\rho}\left[y(\xi')\right]
\right>^{(1)}
+z_\alpha(\xi)z_\beta(\xi')\frac{\partial^2}{\partial 
y_\alpha(\xi)\partial y_\beta(\xi')}\left<B_{\lambda\nu}\left[y(\xi)
\right]B_{\mu\rho}\left[y(\xi')\right]\right>^{(1)}
\Biggr)\Biggr], \eqno (18)$$
where from now on we shall omit all the boundary terms.

Secondly, one should substitute the saddle-point of the integral

$$\int DB_{\mu\nu}\exp\Biggl[-\int dx\Biggl(\frac{1}{12\Lambda^2}
H_{\mu\nu\lambda}^2+\frac{1}{4e^2}B_{\mu\nu}^2\Biggr)+
I^{(0)} \Biggr] $$ 
into Eq. (17). This saddle-point reads 

$$B^{{\rm extr.}}_{\mu\nu}\left[y(\xi)\right]=\frac{ie^2m^3}{2\pi^2}\int 
d\sigma_{\mu\nu}\left(y(\xi')\right)\frac{K_1\left(m\left|y(\xi)-
y(\xi')\right|\right)}{\left|y(\xi)-y(\xi')\right|}, $$
and upon its substitution into Eq. (17), accounting for Eq. (18), and  
making use of Eq. (4), we finally get the following 
value of the action quadratic in quantum 
fluctuations

$$S_{{\rm quadr.}}=\frac{e^2m^3}{4\pi^2}\int d\sigma_{\mu\nu}\left(y(\xi)\right)
\int\frac{d\sigma_{\mu\nu}\left(y(\xi')\right)}{\left|y(\xi)-y(\xi')\right|}
\Biggl\{K_1-\frac{z_\alpha(\xi)(y(\xi)-y(\xi'))_\alpha}{\left|y(\xi)-
y(\xi')\right|}\Biggl(\frac{2K_1}{\left|y(\xi)-y(\xi')\right|}+$$

$$+m\left(K_0+K_2\right)\Biggr)
+\frac{z_\alpha(\xi)z_\beta(\xi')}{\left|y(\xi)-y(\xi')\right|}
\Biggl[\delta_{\alpha\beta}\Biggl(\frac{K_1}{\left|y(\xi)-y(\xi')\right|}+$$

$$+\frac{m}{2}\left(K_0+K_2\right)\Biggr)-\frac{\left(y(\xi)-y(\xi')
\right)_\alpha\left(y(\xi)-y(\xi')\right)_\beta}{\left|y(\xi)-y(\xi')\right|}
\Biggl(3\Biggl(\frac{m^2}{4}+\frac{1}{\left(y(\xi)-y(\xi')\right)^2}\Biggr)
K_1+$$

$$+\frac{3m}{2\left|y(\xi)-y(\xi')\right|}\left(K_0+K_2\right)+\frac{m^2}{4}
K_3\Biggr)\Biggr]\Biggr\}-
\frac{e^2m^3}{2\pi^2}\Biggl(\int d^2\xi z_\nu (\xi)
h_{\nu\mu\lambda}\left[y(\xi)\right]\varepsilon^{ab}\left(\partial_a
y_\mu(\xi)\right)\left(\partial_b z_\lambda(\xi)\right)+$$

$$+\frac{1}{2}
\int d\sigma_{\mu\lambda}\left(y(\xi)\right)z_\nu(\xi)z_\alpha(\xi)
\partial_\alpha h_{\nu\mu\lambda}\left[y(\xi)\right]\Biggr), \eqno (19)$$ 
where 

$$h_{\nu\mu\lambda}\left[y(\xi)\right]\equiv$$

$$\equiv\Biggl[\int d\sigma_{\mu
\lambda}\left(y(\xi')\right)\left(y(\xi)-y(\xi')\right)_\nu+
\int d\sigma_{\nu\mu}\left(y(\xi')\right)\left(y(\xi)-y(\xi')\right)_\lambda
+\int d\sigma_{\lambda\nu}\left(y(\xi')\right)\left(y(\xi)-y(\xi')
\right)_\mu\Biggr]\cdot$$

$$\cdot\frac{1}{\left(y(\xi)-y(\xi')\right)^2}\Biggl[
\frac{K_1}{\left|y(\xi)-y(\xi')\right|}+\frac{m}{2}\left(K_0+K_2\right)
\Biggr], \eqno (20)$$
and everywhere in Eqs. (19) and (20) the arguments of the Macdonald 
functions are the same, $m\left|y(\xi)-y(\xi')\right|$. We have also 
kept the pure background term on the R.H.S. of Eq. (19). The other 
terms yield the couplings of quantum fluctuations with the 
background world-sheet. All of these terms except one do not contain the 
derivatives of quantum fluctuations.

\vspace{6mm}
{\large \bf 4. Conclusion}

\vspace{3mm}
In this Letter, we have modelled the gluodynamics string effective action 
by the theory of the massive Abelian Kalb-Ramond field, 
interacting with the string. 
The partition function (1) of this theory is nothing else but the low-energy 
expression for the partition function of the UCST, which is in fact   
the low-energy limit of the Wilson average in the 4D compact QED, 
rewritten in terms of the dual antisymmetric tensor 
field. This observation provides us with 
a unified description of the gluodynamics string and the UCST. 

An approach to the gluodynamics string with the help of the theory (1) 
has been justified in Section 2, where it has been shown that the small- 
and large-distance  
asymptotic behaviours of the coefficient functions, which parametrize the 
gauge-invariant  
correlator of two gluonic field strength tensors 
within the MVC and could be extracted from the lattice 
measurements, are in a good agreement with the ones of the corresponding 
functions standing in the propagator of the Kalb-Ramond field. 
It has also been demonstrated that the mass of the Kalb-Ramond field 
plays the role of the inverse correlation length of the vacuum, so that 
in the massless limit strings are suppressed, which is in agreement with 
Refs. 4 and 24. 
  
In Section 3, we have applied the background field formalism, developed 
in Ref. 20 for the nonlinear sigma models, in order to describe fluctuations 
of strings in our model. To this end, we have splitted the string 
world-sheet coordinate into a background part, corresponding to some given  
world-sheet (e.g. one with the minimal area), and quantum fluctuations, 
and using the arc-dependent string action as a generating functional, 
performed 
the expansion of the term, which describes the interaction of the 
string with the 
Kalb-Ramond field, up to the second order in quantum fluctuations. Finally, 
upon the integration over the Kalb-Ramond field, we have derived an   
action, which is quadratic in quantum fluctuations and contains the 
pure background part and the terms  
describing the interaction of the background world-sheet with the quantum 
fluctuations. This action is given by formulae (19) and (20) and 
describes fluctuations of strings in the UCST and in our model of the 
gluodynamics string.   

In the forthcoming paper$^{26}$, we shall demonstrate that there in fact 
exists a physical connection between UCST and gluodynamics based on 
the Abelian Higgs Model.

\vspace{6mm}
{\large \bf 5. Acknowledgments}

\vspace{3mm}
I am grateful to Profs. H.G.Dosch, H.Kleinert,  
H.Reinhardt, M.G.Schmidt, and especially to Profs. Yu.M.Makeenko, 
M.I.Polikarpov, and  
Yu.A.Simonov for the useful discussions. I 
would also like 
to thank the theory group of the Quantum Field Theory Department of the 
Institut f\"ur Physik of the Humboldt-Universit\"at of Berlin and 
especially Profs. D.Ebert, D.L\"ust, and M.M\"uller-Preussker   
for kind hospitality.

\vspace{6mm}
{\large \bf Appendix. Integration over the Kalb-Ramond Field in the 
UCST Partition Function (1)}

According to the general rule, in order to calculate 
the Gaussian integral (1) one should substitute its saddle-point value 
back into the integrand. The saddle-point equation in the momentum 
representation reads 

$$\frac{1}{2\Lambda^2}\left(p^2B_{\nu\lambda}^{\rm ext.}(p)+p_\lambda 
p_\mu B_{\mu\nu}^{\rm ext.}(p)+p_\mu p_\nu B_{\lambda\mu}^{\rm ext.}(p)
\right)+\frac{1}{2e^2}B_{\nu\lambda}^{\rm ext.}(p)=iT_{\nu\lambda}(p).$$
This equation can be most easily solved by rewriting it in the 
following way

$$\left(p^2 {\bf P}_{\lambda\nu, \alpha\beta}+m^2 {\bf 1}_{\lambda\nu, 
\alpha\beta}\right)B_{\alpha\beta}^{\rm ext.}(p)=2i\Lambda^2T_{\lambda
\nu}(p), \eqno (A.1)$$
where we have introduced the following projection operators 
(see, for example, 
Appendix to the last Ref. in$^{15}$; functional generalization of these 
operators has been introduced in Ref. 3) 

$${\bf P}_{\mu\nu, \lambda\rho}\equiv\frac12\left({\cal P}_{\mu\lambda}
{\cal P}_{\nu\rho}-{\cal P}_{\mu\rho} {\cal P}_{\nu\lambda}\right)$$
and

$${\bf 1}_{\mu\nu, \lambda\rho}\equiv\frac12 \left(\delta_{\mu\lambda}
\delta_{\nu\rho}-\delta_{\mu\rho}\delta_{\nu\lambda}\right),$$
where ${\cal P}_{\mu\nu}\equiv\delta_{\mu\nu}-\frac{p_\mu p_\nu}{p^2}$. 
These projection operators possess the following properties 

$${\bf 1}_{\mu\nu, \lambda\rho}=-{\bf 1}_{\nu\mu, \lambda\rho}=
-{\bf 1}_{\mu\nu, \rho\lambda}={\bf 1}_{\lambda\rho, \mu\nu}, 
\eqno (A.2)$$

$${\bf 1}_{\mu\nu, \lambda\rho} {\bf 1}_{\lambda\rho, \alpha\beta}=
{\bf 1}_{\mu\nu, \alpha\beta} \eqno (A.3)$$
(the same properties hold for ${\bf P}_{\mu\nu, \lambda\rho}$), and 

$${\bf P}_{\mu\nu, \lambda\rho}\left({\bf 1}-{\bf P}\right)_{\lambda
\rho, \alpha\beta}=0. \eqno (A.4)$$
By virtue of properties (A.2)-(A.4), the solution of Eq. (A.1) reads 

$$B_{\lambda\nu}^{\rm ext.}(p)=\frac{2i\Lambda^2}{p^2+m^2}\left[
{\bf 1}+\frac{p^2}{m^2}\left({\bf 1}-{\bf P}\right)\right]_{\lambda
\nu, \alpha\beta}T_{\alpha\beta}(p),$$
which, once being substituted back into partition function (1), yields 
for it the following expression

$$\left<W(C)\right>=\exp\Biggl[-\Lambda^2\int\frac{dp}{(2\pi)^4}
\frac{1}{p^2+m^2}\left[{\bf 1}+\frac{p^2}{m^2}\left({\bf 1}-{\bf P}
\right)\right]_{\mu\nu, \alpha\beta}T_{\mu\nu}(-p)T_{\alpha\beta}(p)
\Biggr]. 
\eqno (A.5)$$
Rewriting Eq. (A.5) in the coordinate representation we arrive at 
Eq. (2). From Eq. (A.5) one can easily see that the term on its R.H.S. 
proportional to the projection operator ${\bf 1}-{\bf P}$ yields in the 
coordinate representation the 
boundary term, i.e. Eq. (5), whereas 
the term proportional to the projection operator ${\bf 1}$ yields 
Eq. (4).

\vspace{6mm}
{\large \bf References}

\vspace{3mm}
\noindent
1.~A.M.Polyakov, {\it Nucl.Phys.} {\bf B486}, 23 (1997) 
({\it hep-th}/9607049).\\
2.~Yu.M.Makeenko and A.A.Migdal, {\it Phys.Lett.} {\bf B88}, 135 (1979) 
({\it Erratum-ibid.} {\bf B89}, 437 (1980)); A.M.Polyakov, {\it Nucl.Phys.} 
{\bf B164}, 171 (1980); Yu.M.Makeenko and A.A.Migdal, {\it Nucl.Phys.} 
{\bf B188}, 269 (1981); 
for a review see A.A.Migdal, {\it Phys.Rep.} {\bf 102}, 199 (1983).\\
3.~D.V.Antonov, {\it Mod.Phys.Lett.} {\bf A12}, 1419 (1997) 
({\it hep-th}/9703050).\\
4.~M.C.Diamantini, F.Quevedo, and C.A.Trugenberger, {\it Phys.Lett.} 
{\bf B396}, 115 (1997).\\
5.~H.Kleinert, {\it Phys.Lett.} {\bf B211}, 151 (1988); K.I.Maeda and 
N.Turok, {\it Phys.Lett.} {\bf B202}, 376 (1988); S.M.Barr and D.Hochberg, 
{\it Phys.Rev.} {\bf D39}, 2308 (1989); H.Kleinert, {\it Phys.Lett.} 
{\bf B246}, 127 (1990); P.Orland,{\it Nucl.Phys.} {\bf B428}, 221 (1994); 
H.Kleinert and 
A.M.Chervyakov, {\it hep-th}/9601030 (in press in {\it Phys.Lett.} 
{\bf B}); M.Anderson, F.Bonjour, R.Gregory, and J.Stewart, {\it 
hep-ph}/9707324.\\
6.~A.M.Polyakov, {\it Gauge Fields and Strings} (Harwood Academic 
Publishers, 1987).\\
7.~D.V.Antonov, D.Ebert, and Yu.A.Simonov, {\it Mod.Phys.Lett.} {\bf
A11}, 1905 (1996) (preprint DESY 96-134) ({\it hep-th}/9605086).\\
8.~D.V.Antonov, {\it JETP Lett.} 
{\bf 65}, 701 (1997) ({\it hep-th}/9612109).\\
9.~D.V.Antonov, {\it hep-th}/9705073; D.V.Antonov and D.Ebert, 
{\it Mod.Phys.Lett.} {\bf A12}, 2047 (1997) 
(preprint HU-EP-97/41) ({\it hep-th}/9707097).\\
10. H.G.Dosch, {\it Phys.Lett.} {\bf B190}, 177 (1987); Yu.A.Simonov,
{\it Nucl.Phys.} {\bf B307}, 512 (1988); H.G.Dosch and Yu.A.Simonov,
{\it Phys.Lett.} {\bf B205}, 339 (1988), {\it Z.Phys.} {\bf C45}, 147
(1989); Yu.A.Simonov, {\it Nucl.Phys.} {\bf B324}, 67 (1989), {\it
Phys.Lett.} {\bf B226}, 151 (1989), {\it Phys.Lett.} {\bf B228}, 413
(1989), {\it Yad.Fiz.} {\bf 54}, 192 (1991).\\
11. A.Yu.Dubin, A.B.Kaidalov, and Yu.A.Simonov, {\it Phys.Lett.} 
{\bf B323}, 41 (1994); E.L.Gubankova and A.Yu.Dubin, {\it Phys.Lett.} 
{\bf B334}, 180 (1994); E.L.Gubankova and Yu.A.Simonov, {\it Phys.Lett.} 
{\bf B360}, 93 (1995).\\ 
12. Yu.A.Simonov, {\it Nuovo Cim.} {\bf A107}, 2629 (1994).\\
13. D.V.Antonov and Yu.A.Simonov, {\it Int.J.Mod.Phys.} {\bf A11}, 4401 
(1996).\\
14. D.V.Antonov, {\it Int.J.Mod.Phys.} {\bf A12}, 2047 (1997) 
({\it hep-th}/9606195).\\
15. D.V.Antonov, {\it JETP Lett.} {\bf 63}, 398 (1996) 
({\it hep-th}/9605031), 
{\it Phys.Atom.Nucl.} {\bf 60}, 299, 478 (1997) ({\it hep-th}/9605044, 
{\it hep-th}/9605045).\\ 
16. Yu.A.Simonov and V.I.Shevchenko, {\it Phys.Atom.Nucl.} {\bf 60}, 1201 
(1997) ({\it hep-th}/9701026).\\
17. Yu.A.Simonov, {\it Yad.Fiz.} {\bf 50}, 213 (1989).\\
18. A. Di Giacomo and H.Panagopoulos, {\it Phys.Lett.} {\bf B285}, 133 
(1992); A. Di Giacomo, E.Meggiolaro, and H.Panagopoulos, 
{\it hep-lat}/9603017 (preprints IFUP-TH 12/96 and UCY-PHY-96/5) 
(in press in {\it Nucl.Phys.} {\bf B}).\\ 
19. M. D'Elia, A. Di Giacomo, and E.Meggiolaro, {\it hep-lat}/9705032.\\
20. E.Braaten, T.L.Curtright, and C.K.Zachos, {\it Nucl.Phys.} 
{\bf B260}, 630 (1985).\\
21. Yu.A.Simonov, {\it Yad.Fiz.} {\bf 58}, 113 (1995).\\
22. M.Eidem\"uller and M.Jamin, {\it hep-ph}/9709419; Yu.A.Simonov and 
V.I.Shevchenko, in preparation.\\ 
23. H.G.Dosch, privat communication.\\
24. Yu.A.Simonov, preprint ITEP-PH-97-4 ({\it hep-ph}/9704301).\\
25. M.I.Polikarpov, U.-J.Wiese, and M.A.Zubkov, {\it Phys.Lett.} 
{\bf B309}, 133 (1993); P.Orland, {\it Nucl.Phys.} {\bf B428}, 
221 (1994), M.Sato and S.Yahikozawa, {\it Nucl.Phys.} {\bf B436}, 
100 (1995); 
E.T.Akhmedov, M.N.Chernodub, M.I.Polikarpov, and M.A.Zubkov, 
{\it Phys.Rev.} {\bf D53}, 2087 (1996); E.T.Akhmedov, {\it JETP Lett.} 
{\bf 64}, 82 (1996).\\
26. D.V.Antonov, in preparation.

\end{document}